\documentclass[aps,prb,preprint,groupedaddress,showpacs,amsmath,amssymb]{revtex4}

\usepackage{graphicx}% Include figure files
\usepackage{dcolumn}% Align table columns on decimal point
\usepackage{bm}% bold math

\begin{document}

%Title of paper
\title{Effect of pressure on the structural phase transition and superconductivity in (Ba$_{1-x}$K$_x$)Fe$_2$As$_2$ ($x = 0$ and $0.45$) single crystals.}

\author{M. S. Torikachvili}
\affiliation{Department of Physics, San Diego State University, San Diego, CA 92182-1233}
\author{N. Ni}
\author{S. L. Bud'ko}
\author{P. C. Canfield}
\affiliation{Ames Laboratory US DOE and Department of Physics and Astronomy, Iowa State University, Ames, Iowa 50011}

\date{\today}

\begin{abstract}
The effects of pressure up to $\sim 20$ kbar, on the structural phase transition of lightly Sn-doped BaFe$_2$As$_2$, as well as on the superconducting transition temperature and upper critical field of (Ba$_{0.55}$K$_{0.45}$)Fe$_2$As$_2$ single crystals have been studied. Both transitions decrease with pressure in an almost linear fashion. Under pressure, the upper critical field curve, $H_{c2}(T)$, shifts down in temperature to follow the zero field $T_c$ with very little change in slope.
\end{abstract}

% insert suggested PACS numbers in braces on next line
% insert suggested keywords - APS authors don't need to do this
%\keywords{}

%\maketitle must follow title, authors, abstract, \pacs, and \keywords
\maketitle
The recent reports of the discovery of elevated temperature superconductivity in LaFeAs(O$_{1-x}$F$_{x}$) \cite{jap_dis} followed by an almost two-fold increase in the superconducting transition temperature by application of pressure \cite{jap_pre} or by substitution of heavier rare earths \cite{fst_sm} has brought a lot of attention to materials with structures containing Fe-As layers. Within a few months, superconductivity below $T_c = 38$ K was discovered in the structurally related, non-oxide material, K-doped BaFe$_2$As$_2$ \cite{sec_ger} and single crystals of the parent, non-superconducting compound, BaFe$_2$As$_2$, as well as superconducting (Ba$_{0.55}$K$_{0.45}$)Fe$_2$As$_2$ were synthesized. \cite{ni_ba,ba_cn} Although the parent compound appears to be sensitive to small, $\sim 1$\%, of Sn doping, \cite{ni_ba,ba_cn,fst_ger} a small amount of Sn does not seem to be detrimental for superconductivity in the K-doped compound (light doping of Sn is an unintended consequence of the crystal growth process). \cite{sec_ger,ni_ba}  Bearing in mind the significant effect of pressure on the structural phase transition and superconductivity in the related, RFeAs(O$_{1-x}$F$_{x}$) (R = rare earth), family of materials, in this work we study the effect of hydrostatic pressure on parent, lightly Sn-doped, BaFe$_2$As$_2$, as well as superconducting (Ba$_{0.55}$K$_{0.45}$)Fe$_2$As$_2$ single crystals.
\\

Single crystals of BaFe$_2$As$_2$ and (Ba$_{0.55}$K$_{0.45}$)Fe$_2$As$_2$ were grown out of a Sn flux, using conventional high temperature solution growth techniques. \cite{can_fi}  The details of the growth as well as thermodynamic and transport properties of these crystals are described in Ref. \onlinecite{ni_ba}. At ambient pressure the structural phase transition in BaFe$_2$As$_2$ is marked by a rapid increase of in-plane resistivity.
The pressure dependencies of the structural phase transition temperature, $T_0$, the superconducting phase transition temperature, $T_c$, and the upper critical field, $H_{c2}$, were determined from the temperature-dependent in-plane resistance. Pressure was generated in a Teflon cup filled with either a 60:40 mixture of n-pentane and light mineral oil (BaFe$_2$As$_2$) or Fluorinert FC-75 ((Ba$_{0.55}$K$_{0.45}$)Fe$_2$As$_2$) inserted in a 22 mm outer diameter, non-magnetic, piston-cylinder-type, Be-Cu pressure cell with a core made of NiCrAl (40 KhNYu-VI) alloy. The pressure at room temperature was monitored by a manganin, resistive pressure gauge. At low temperatures the pressure value was determined from the superconducting transition temperature of pure lead \cite{eil81a}. Low temperature pressure values will be used throughout the text as the pressure remains almost constant in similar geometry cells on cooling below $\sim 100$ K. \cite{joe} The temperature and magnetic field environment for the pressure cell was provided by a Quantum Design Physical Property Measurement System (PPMS-9) instrument. An additional Cernox sensor, attached to the body of the cell, served to determine the temperature of the sample for these measurements. The cooling rate was below 0.5 K/min, the temperature lag between the Cernox on the body of the cell and the system thermometer was $< 0.5$ K at high temperatures and 0.1 K or less below $\sim 70$ K.
\\

\begin{figure}
\begin{center}
\includegraphics[angle=0,width=90mm]{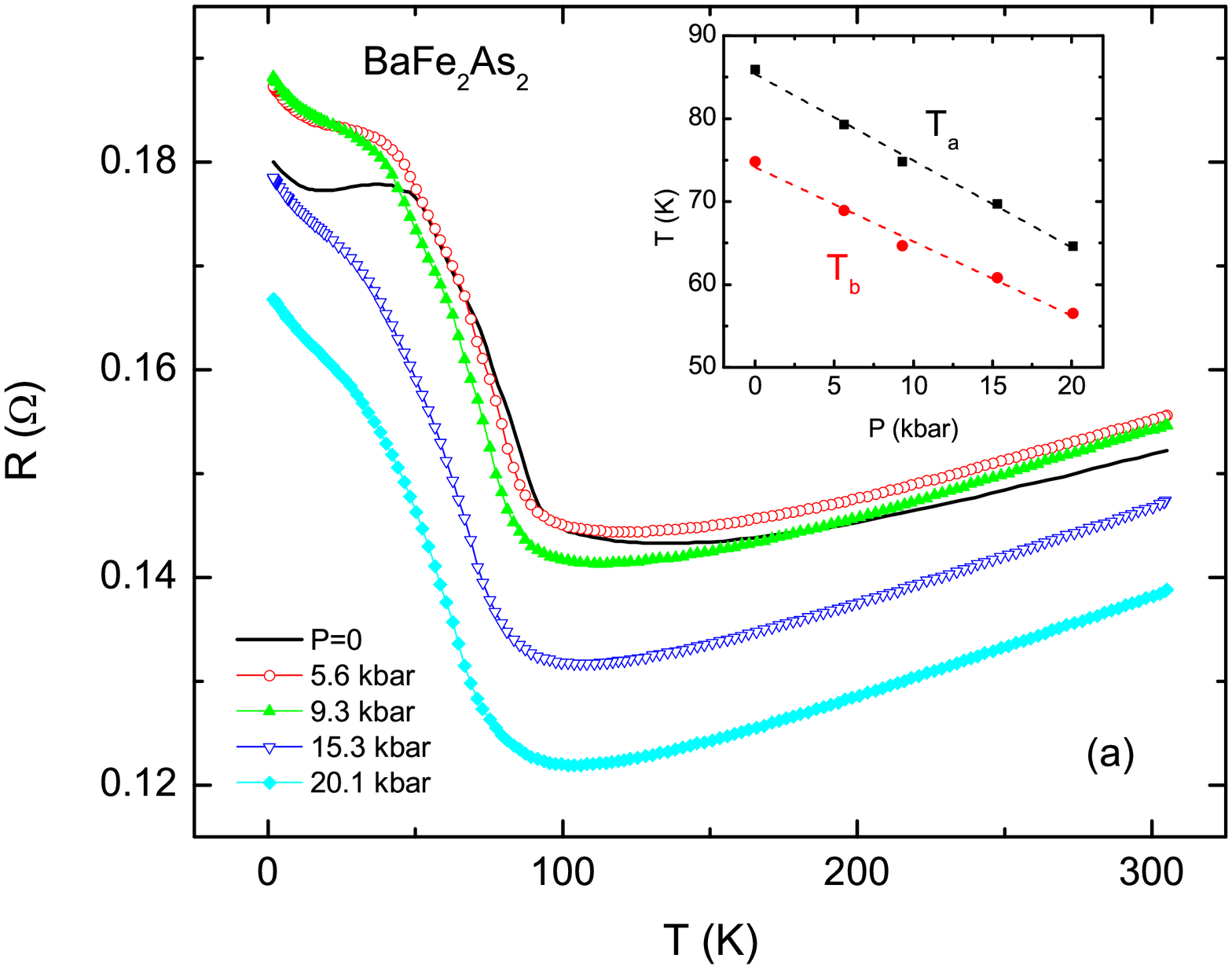}
\includegraphics[angle=0,width=90mm]{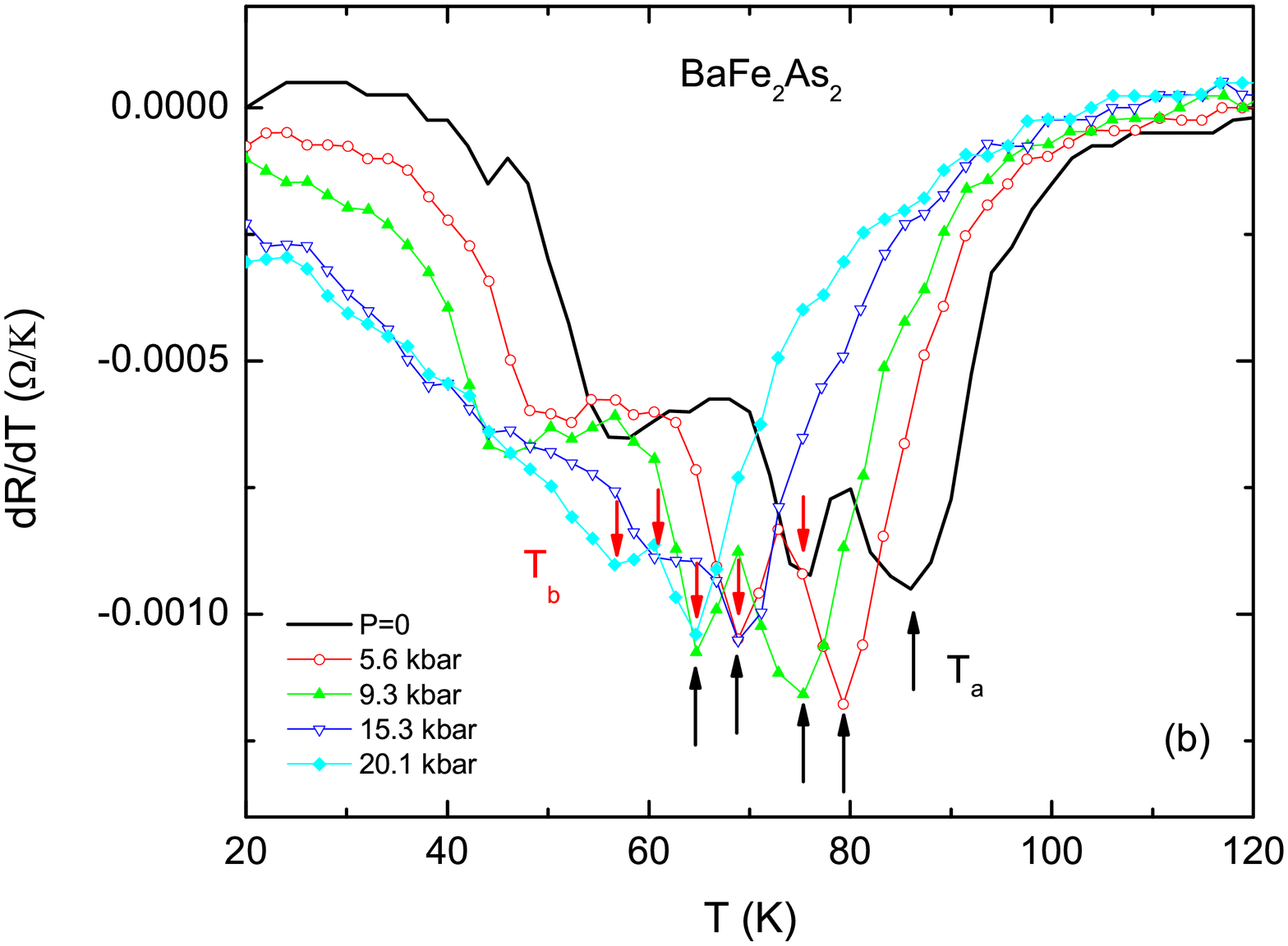}
\end{center}
\caption{(Color online) (a)Pressure dependence of in-plane resistance of BaFe$_2$As$_2$. Inset: pressure dependent transition temperatures, determined as shown in panel (b), the lines are from linear fits. (b)Derivatives, $dR/dT$, at different pressures in the transition region. Arrows show two definitions of the transition temperature.}\label{F1}
\end{figure}

Figure \ref{F1}(a) shows the temperature dependent resistivity of BaFe$_2$As$_2$ at different pressures. Above $\sim 10$ kbar the $\rho(T)$ curves shift down with pressure. Although the feature in resistivity corresponding to the structural phase transition is somewhat broad, its pressure dependence can be monitored by following the pressure evolution of the minima in the derivative, $dR/dT$ (Fig. \ref{F1}(b)). It is noteworthy, that at least two minima, can be seen in $dR/dT$ (marked with up- ($T_a$) and down- ($T_b$) arrows in Fig. \ref{F1}(b)). Both minima shift to lower temperatures (inset to Fig. \ref{F1}(a)) under pressure with similar pressure derivatives, $dT_a/dP = -1.04 \pm 0.04$ K/kbar, $dT_b/dP = -0.89 \pm 0.05$ K/kbar. Using a linear extrapolation of these data, the structural phase transition can be expected to be suppressed by $\sim 80$ kbar. This is most likely an upper limit given the possibility of super-linear suppression at higher pressures.
\\

\begin{figure}
\begin{center}
\includegraphics[angle=0,width=90mm]{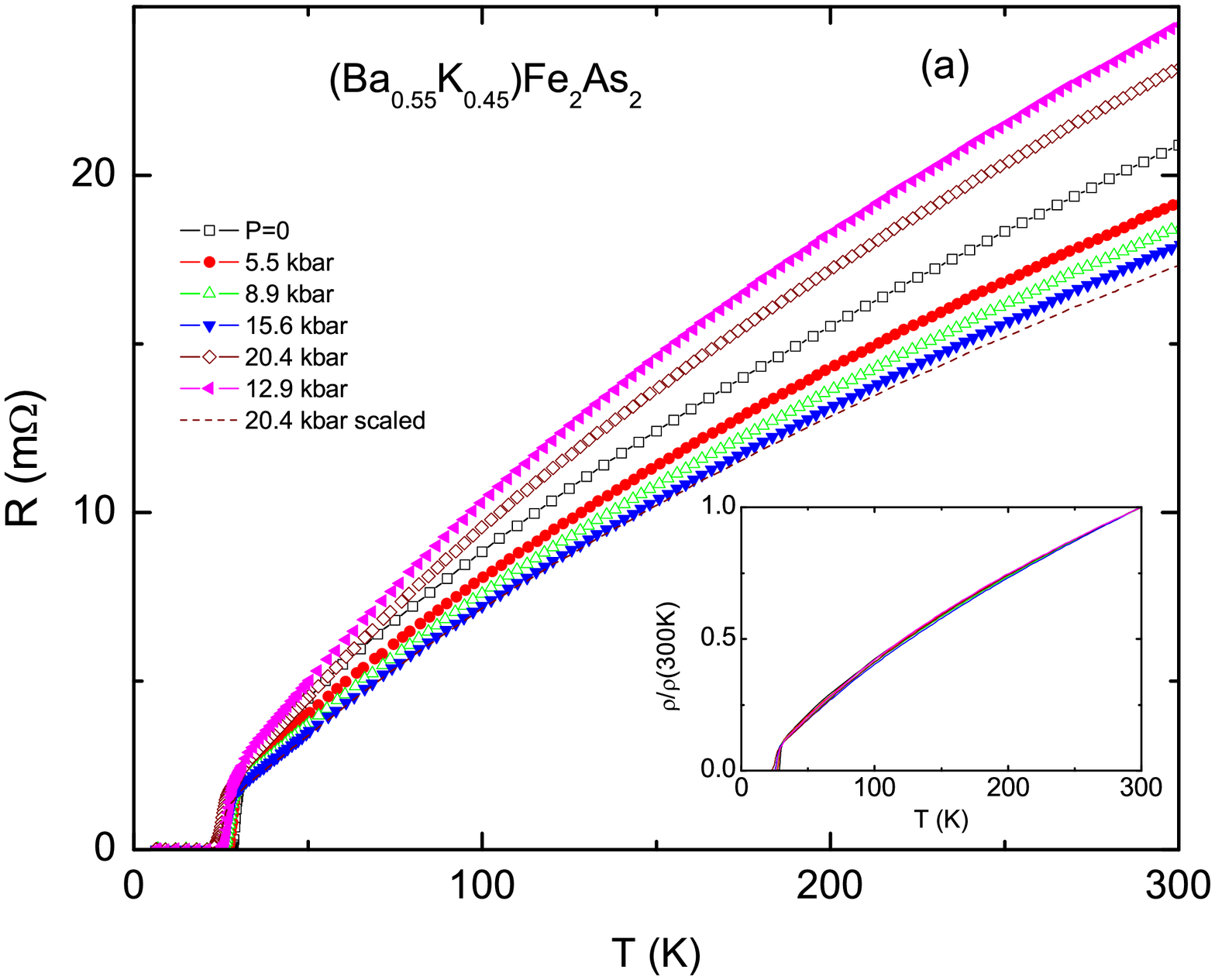}
\includegraphics[angle=0,width=90mm]{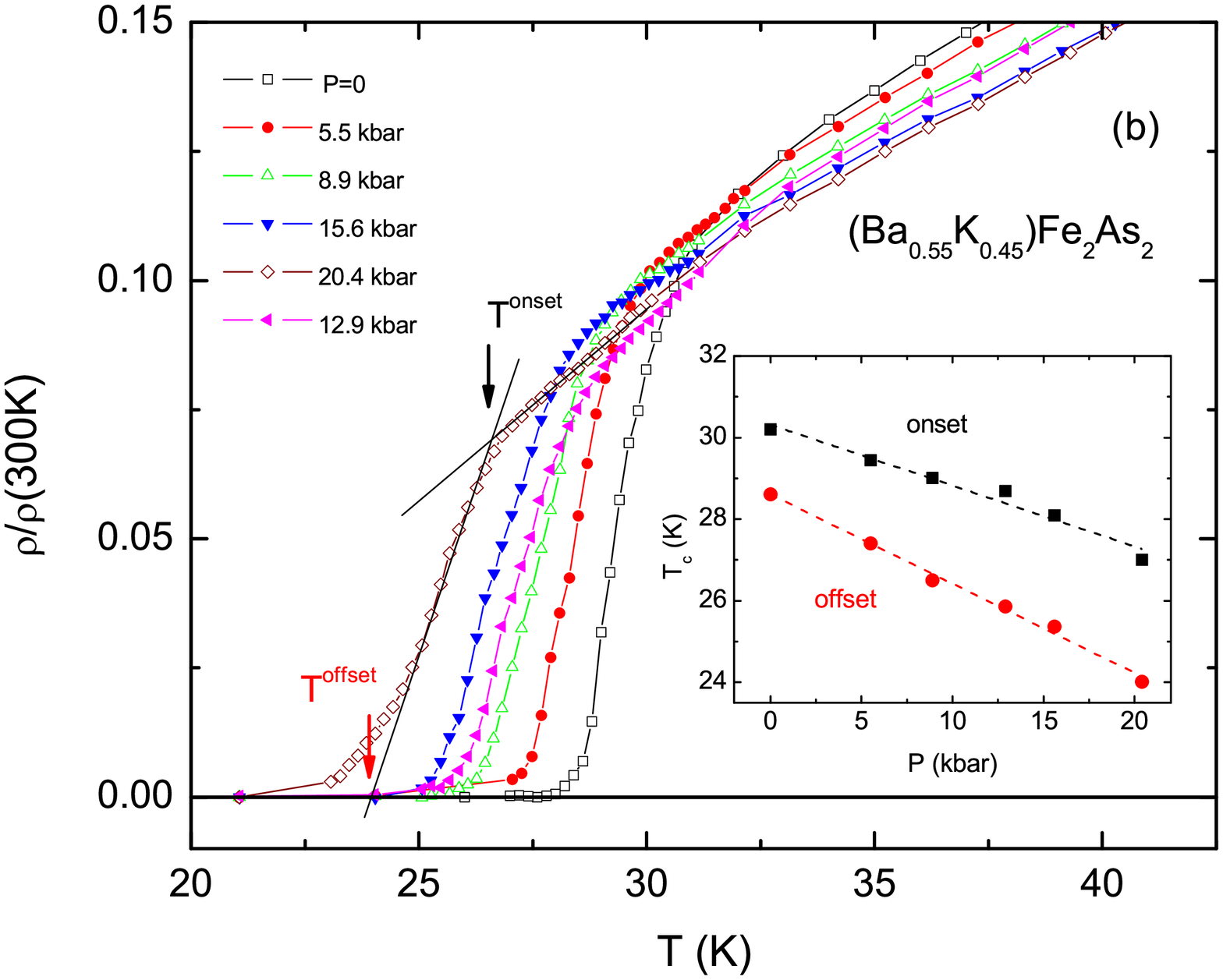}
\end{center}
\caption{(Color online) (a) Temperature dependent in-plane resistance of (Ba$_{0.55}$K$_{0.45}$)Fe$_2$As$_2$ under pressure (pressure values in the legend are in the order of runs). Dashed line - 20.4 kbar data scaled in a way that brings the last, 12.9 kbar, data between 8.9 kbar and 15.6 kbar runs. Inset: normalized temperature dependent resistivity as a function of pressure. (b) Resistive superconducting transition in (Ba$_{0.55}$K$_{0.45}$)Fe$_2$As$_2$ (pressure values in the legend are in the order of runs). Inset: superconducting transition temperatures (defined as onset and offset of resistive transition) as a function of pressure. The lines are from linear fits.}\label{F2}
\end{figure}

For superconducting (Ba$_{0.55}$K$_{0.45}$)Fe$_2$As$_2$, the normal state resistivity decreases under pressure up to $\sim 20$ kbar as shown in Fig. \ref{F2}(a).  However the normalized resistivity, $\rho(T)/\rho(300K)$ does not change significantly (Fig. \ref{F2}(a), inset). (Note that apparently the sample has developed a small crack after the 4th pressure run, 15.6 kbar, that caused an increase in the measured resistance for the next two runs, however the 20.4 kbar run yields  consistent values if scaled with the last, 12.9 kbar, curve, as shown in Fig. \ref{F2}(a)) The superconducting transition temperature decreases under pressure with some (reversible) broadening of the resistive transition (Fig. \ref{F2}(b)). For different criteria in the determination of $T_c$, the pressure derivatives are  $dT_c^{onset}/dP = -0.15 \pm 0.01$ K/kbar, $dT_c^{offset}/dP = -0.21 \pm 0.01$ K/kbar (Fig. \ref{F2}(b)).

\begin{figure}
\begin{center}
\includegraphics[angle=0,width=120mm]{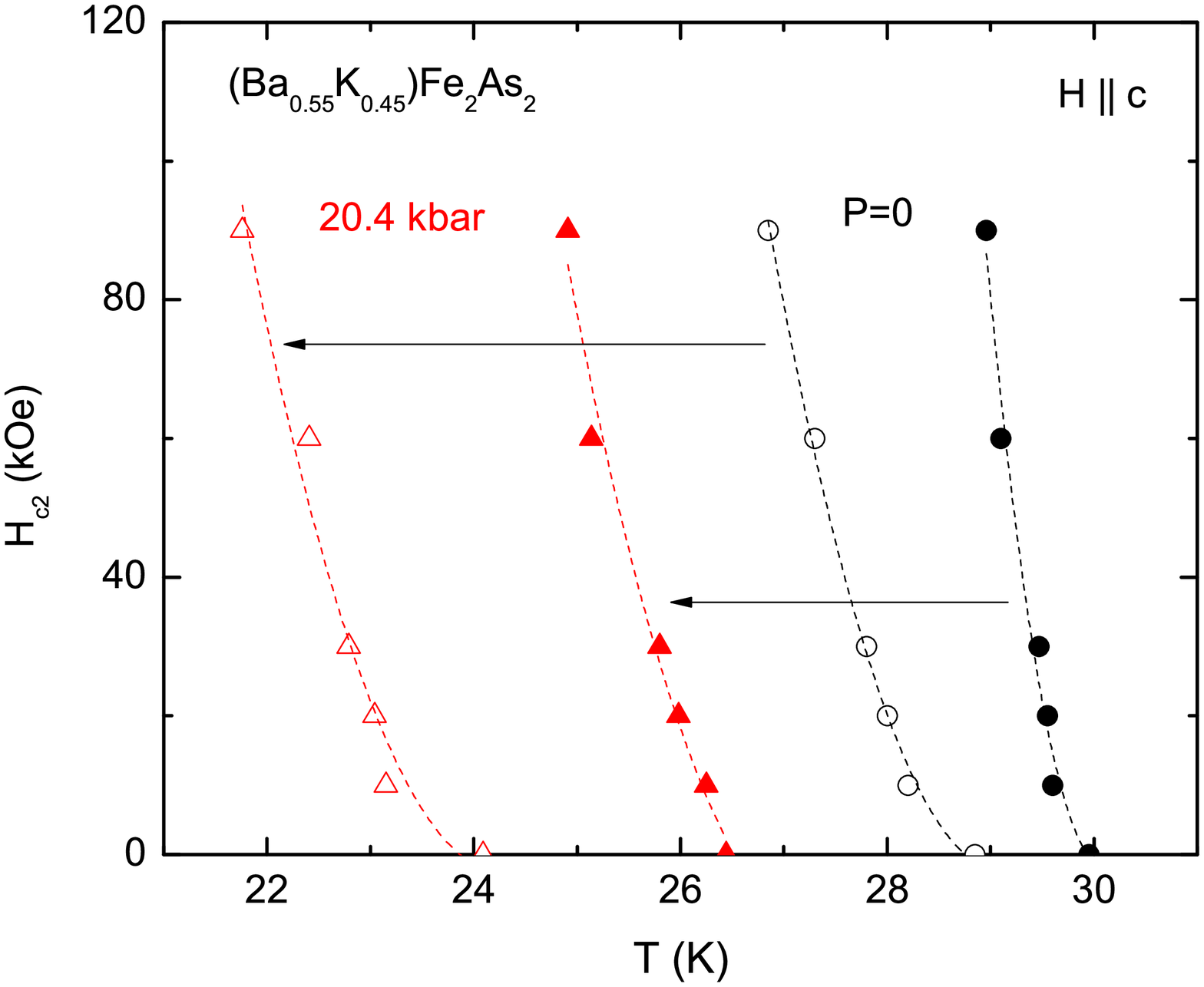}
\end{center}
\caption{(Color online) Upper critical fields, $H_{c2}(T)$ measured up to 90 kOe at zero and maximum pressure as defined from onset (filled symbols) and offset (open symbols) of resistive transitions. The lines are guides to the eye.}\label{F3}
\end{figure}

The upper critical field in (Ba$_{0.55}$K$_{0.45}$)Fe$_2$As$_2$ is expected to be extremely high, on the order of 1,000 kOe \cite{ni_ba}. Our measurements, up to 90 kOe, can probe only a small section of the $H_{c2}(T)$ curve, close to zero-field $T_c$. Under pressure, the $H_{c2}(T)$ appear to shift (Fig \ref{F3}) following the shift of $T_{c0}$ without changes in the slope or curvature.
\\

{\it To summarize}, in lightly Sn-doped BaFe$_2$As$_2$ the structural phase transition is suppressed by application of pressure approximately two times faster than in non-superconducting SmFeAs(O$_{0.95}$F$_{0.05}$) \cite{uh_sm}. A moderate pressure of $\sim 80$ kbar, or less, is expected to suppress the structural phase transition completely.

For superconducting RFeAs(O$_{1-x}$F$_x$) a variety of pressure dependencies have been reported: initial increase of $T_c$, followed by a maximum and almost decrease with pressure was reported for LaFeAs(O$_{0.89}$F$_{0.11}$) \cite{jap_pre,zoc_P}, non-linear $T_c$ increase (LaFeAs(O$_{0.95}$F$_{0.05}$) \cite{jap_pre}, SmFeAs(O$_{0.87}$F$_{0.13}$) \cite{uh_sm}) or decrease (CeFeAs(O$_{0.88}$F$_{0.12}$) \cite{zoc_P}), as well as close-to-linear pressure dependencies of different signs and values (RFeAs(O$_{1-x}$F$_x$) \cite{uh_sm,tak_jacs}). The negative, rather large, pressure derivative of $T_c$ observed in (Ba$_{0.55}$K$_{0.45}$)Fe$_2$As$_2$ is well within the range of the available data for oxygen-containing RFeAs(O$_{1-x}$F$_x$). It is possible, if the superconducting phase diagram in (Ba$_{1-x}$K$_{x}$)Fe$_2$As$_2$ has a dome-like shape as a function of K-concentration and pressure, that (Ba$_{0.55}$K$_{0.45}$)Fe$_2$As$_2$ is positioned in the slightly overdoped region.

The results above suggest several extensions: (i)search for superconductivity in BaFe$_2$As$_2$ at higher pressures, perhaps up to $\sim 80$ kbar; (ii)detailed study of pressure dependencies in K-doped materials with several values of K-concentrations, if these samples can be reproducibly grown in single crystal form; (iii)high field studies in $H_{c2}(T)$ under pressure.

\begin{acknowledgments}
Work at the Ames Laboratory was supported by the US Department of Energy - Basic Energy Sciences under Contract No. DE-AC02-07CH11358.  MST gratefully acknowledges support of the National Science Foundation under DMR-0306165 and DMR-0805335. SLB thanks Starik Kozlodoyev for relevant insights.
\end{acknowledgments}

\clearpage

\clearpage

\clearpage

\begin{thebibliography}{99}

\bibitem{jap_dis}  Y. Kamihara, T. Watanabe, M. Hirano, H Hosono, Journal of the American Chemical Society  {\bf 130}, 3296 (2008).

\bibitem{jap_pre}  H. Takahashi, K. Igawa, K. Arii, Y. Kamihara, M. Hirano, H. Hosono, Nature (London)  {\bf 453}, 376 (2008).

\bibitem{fst_sm}  X. H. Chen, T. Wu, G. Wu, R. H. Liu, H. Chen, D. F. Fang, Nature (London)  {\bf 453}, 761 (2008).

\bibitem{sec_ger}  M. Rotter, M. Tegel, D. Johrendt, arXiv:0805.4630, unpublished.

\bibitem{ni_ba} N. Ni, S. L. Bud'ko, A. Kreyssig, S. Nandi, G. E. Rustan, A. I. Goldman, S. Gupta, J. D. Corbett, A. Kracher, and P. C. Canfield, arXiv:0806.1874, unpublished.

\bibitem{ba_cn} X. F. Wang, T. Wu, G. Wu, H. Chen, Y. L. Xie, J. J. Ying, Y. J. Yan, R. H. Liu, X. H. Chen, arXiv:0806.2452, unpublished.

\bibitem{fst_ger}  M. Rotter, M. Tegel, D. Johrendt, I. Schellenberg, W. Hermes, R. Poettgen, arXiv:0805.4021, unpublished.

\bibitem{can_fi}  P. C. Canfield, Z.  Fisk, Phil. Mag. B  {\bf 65}, 1117 (1992).

\bibitem{eil81a} A. Eiling, J. S. Schilling, J. Phys. F: Met. Phys. {\bf 11}, 623 (1981).

\bibitem{joe} J. D. Thompson, Rev. Sci. Instrum., {\bf 55}, 231 (1984).

\bibitem{uh_sm} B. Lorenz, K. Sasmal, R. P. Chaudhury, X. H. Chen, R. H. Liu, T. Wu, C. W. Chu, arXiv:0804.1582, unpublished.

\bibitem{zoc_P} D. A. Zocco, J. J. Hamlin, R. E. Baumbach, M. B. Maple, M. A. McGuire, A. S. Sefat, B.C. Sales, R. Jin, D. Mandrus, J.B. Jeffries, S. T. Weir, Y. K. Vohra, arXiv:0805.4372, unpublished.

\bibitem{tak_jacs}  Y. Takabayashi, M. T. McDonald, D. Papanikolaou, S. Margadonna, G. Wu, R. H. Liu, X. H. Chen, K. Prassides, J. Amer. Chem. Soc. {\bf 130} (2008), in press.

\end{thebibliography}
\end{document}